\documentclass[twocolumn,shortnote]{jpsj2}
\usepackage{graphicx}

\def\runtitle{Kondo effect in two-dimensional disordered systems}
\def\runauthor{Sei-ichiro \textsc{Suga} and Takuma \textsc{Ohashi}}

\title{%
Kondo Effect in Two-Dimensional Disordered Electron Systems
}

\author{%
Sei-ichiro \textsc{Suga}\thanks{suga@tp.ap.eng.osaka-u.ac.jp} and
Takuma \textsc{Ohashi}
}

\inst{%
Department of Applied Physics, Osaka University, Suita, Osaka 565-0871 
}

\recdate{\today}

\abst
{
}

\kword{%
 Kondo effect, Anderson localization, non-Fermi liquid, distribution of Kondo temperature
}

\begin{document}
\sloppy
\maketitle

In disordered systems, the localization effect of the conduction electron enhances the electron correlation effects drastically, leading to conspicuous features particularly in low-dimensional systems \cite{AA,HF,LR,Fink,DiC,BK}.
The Kondo effect is a typical phenomenon caused by the electron correlation around a magnetic impurity. Using perturbative expansion, the Kondo effect in the weakly localized regime was investigated \cite{Ohkawa}. It was shown that the Kondo logarithmic terms are modified into the product of new anomalous terms and the Kondo logarithmic ones, and that the latters are scaled into the same Kondo temperature as that without randomness \cite{Suga}. 
The Kondo effect in strongly disordered systems was studied by taking account of the Coulomb interaction among conduction electrons \cite{DKK}.  It was shown that the Kondo temperature has a spatial distribution, which leads to divergence behaviors of physical quantities as temperature approaches zero. 
Such a spatial distribution in the Kondo temperature was suggested from experimental results of strong broadening of the Cu NMR line of ${\rm UCu_{5-x}Pd_x}$ \cite{NMR}. 
In spite of these findings, the effects of strong randomness itself on the behavior of a magnetic impurity have not yet been fully investigated from a microscopic viewpoint.

In this paper, we study the Kondo effect in two-dimensional (2D) strongly disordered electron systems using a finite-temperature quantum Monte Carlo (QMC) method \cite{Hirsch}. 
Let us consider the single-impurity Anderson model with on-site random potentials described by the Hamiltonian 
\begin{eqnarray}
	H 	&=&	\sum_{i \sigma} \epsilon_{i} c_{i\sigma}^{\dag}c_{i\sigma} 
			- t \sum_{<ij> \sigma} c_{i\sigma}^{\dag}c_{j\sigma} 
			+ \epsilon_{d} \sum_{\sigma}n_{d \sigma} 
													\nonumber \\
		& & + V\sum_{\sigma}(d_{\sigma}^{\dag}c_{0\sigma} + H.c.) 
			+ Un_{d\uparrow}n_{d\downarrow}
\end{eqnarray}
where random on-site potentials $\epsilon_{i}$ are chosen to be a flat distribution in the interval $[-W, W]$ under the condition $\sum_{i} \epsilon_{i} = 0$, 
$<i,j>$ denotes the summation of the nearest-neighbor sites, and $n_{d \sigma}=d_{\sigma}^{\dag}d_{\sigma}$. 
The system consists of a $41 \times 41$ square lattice with a magnetic impurity. For $W \geq 3.0$, the conduction electron is localized with the localization length $\xi(W) \leq 37.5$ \cite{MacKinnon}. Thus, the system for $W \geq 3.0$ is probably in the strongly localized regime in low temperature. 
We set the condition $\epsilon_{d}+(1/2)U=0$ and use the parameters $U=2.0$ and $V=-1.0$ in units of $t$. 
The Kondo temperature without randomness can be estimated as \cite{TW} $T_K^0 \sim 0.14$.

%
\begin{figure}[h]
\includegraphics[trim=2cm 5cm 0cm 4cm,clip,width=8cm]{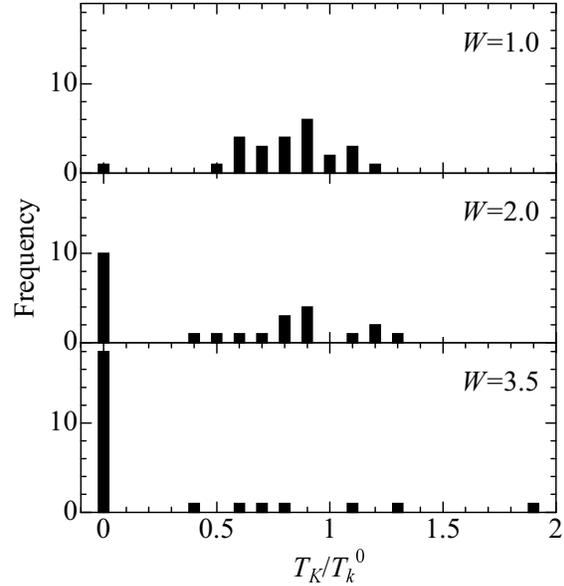}
\vspace{-0.25cm}
\caption{Distribution of the Kondo temperature for $W=1.0$, $W=2.0$ and $3.5$. Twenty-five positions of a magnetic impurity is used around the center. 
}
\label{Fig1}
\end{figure}
%

Shifting the position of a magnetic impurity twenty-four times around the center in the same realization of the random potential, we calculate temperature dependence of the susceptibility of a magnetic impurity $\chi(T)$ in $0 \leq W \leq 3.5$. Depending on the position of a magnetic impurity in given $W$, its local moment can be screened or unscreened by the spin of the conduction electron. In the former case $\chi(T)$ shows a local Fermi-liquid behavior, while in the latter case $\chi(T)$ shows power-law or logarithmic divergence. 
For $W=3.5$, $\chi(T)$ shows a local Fermi-liquid behavior at seven positions. At three positions $\chi(T)$ shows logarithmic divergence, at nine positions $\chi(T)$ shows weak power-law divergence, and at the rest six positions $\chi(T)$ behaves as a free spin; $\chi(T) \sim T^{-1}$. 
The exponents of the weak power-law divergence for $W=3.5$ are summarized in Table I.

The relation between the behaviors of $\chi(T)$ and the local density of states for the conduction electron at the Fermi energy $\rho^i_0$ are summarized in Table II for $W=3.5$. Note that $\rho^i_0$ is obtained by numerical diagonalization method in the case of $U=V=0$. 
As shown in Table II, $0.049 \leq \rho^i_0 \leq 0.549$ for a local Fermi-liquid behavior, $0.105 \leq \rho^i_0 \leq 0.146$ for logarithmic divergence, $0.043 \leq \rho^i_0 \leq 0.162$ for weak power-law divergence, and $0.012 \leq \rho^i_0 \leq 0.032$ for a free spin. 
Therefore, a smaller $\rho^i_0$ has a tendency to make the localized spin unscreened, yielding stronger divergence of $\chi(T)$ with decrease of $T$. 
%
%
\begin{fulltable}
\caption{
Weak divergence exponents of $\chi(T) \sim T^{-\alpha}$. }
\begin{tabular}{cccccccccc} \hline 
Result number & 
$\# 1$ & 
$\# 2$ & 
$\# 3$ & 
$\# 4$ & 
$\# 5$ & 
$\# 6$ & 
$\# 7$ & 
$\# 8$ & 
$\# 9$ \\ \hline
$\alpha$ & 
0.952 & 
0.938 & 
0.807 & 
0.751 & 
0.597 & 
0.485 & 
0.485 & 
0.414 & 
0.343 \\
\hline 
%
\end{tabular}
\end{fulltable}

The results in Table II also indicate that the Kondo temperature $T_K$ has a spatial distribution down to $T_K =0$, since $T_K =[2\pi\chi (0)]^{-1}$ \cite{TW}. We extrapolate $\chi (0)$ and evaluate the distribution of the Kondo temperature. The results are summarized in Fig. 1. As $W$ increases, the distribution of the Kondo temperature becomes wider and the weight at $T_K =0$ increases considerably. We have thus shown that the spatial distribution of the Kondo temperature can be caused only by the effects of a random potential.

To obtain the local information on Kondo screening, we calculate the correlation function between the local moment and the spin of the conduction electron. When the local moment is unscreened, the antiferromagnetic correlation at the magnetic impurity is suppressed noticeably in contrast with the results for the screened case \cite{OS}. 
This suppression of the antiferromagnetic correlation at the magnetic impurity probably cause incomplete screening and a divergence behavior of the susceptibility.

In the disordered systems with dilute magnetic impurities, where each magnetic impurity acts as a single magnetic impurity, the observable susceptibility may be obtained by averaging over the susceptibility at each position of a magnetic impurity. We thus take an average over the susceptibilities at twenty-five positions around the center of the system. The results are shown in Fig. 2. 
For $W=1.5$, the average susceptibility $\chi_{\rm av}(T)$ shows logarithmic divergence in  $0.05 \leq T < T_K^0 \sim 0.14$. 
For $W=2.0$ and $3.5$, $\chi_{\rm av}(T)$ shows weak power-law divergence 
%
%
%
%
\begin{table}
\caption{
Relation between $\chi(T)$ and ${\rho^i_0}$ for $W=3.5$ at each shifted position of a magnetic impurity. The numbers in the top row and the column of the left-hand side represent the coordinates along the horizontal and vertical axes, respectively. The position (21, 21) is the center of the system. FL, log-D and FS denote local Fermi-liquid behavior, logarithmic divergence and behavior of a free spin, respectively. $\# 1-\# 9$ correspond to the results for the weak power-law divergence shown in Table I.   }
%
\begin{tabular}{cccccc}
\hline
& 19 & 20 & 21 & 22 & 23 \\
\hline
19 & FL	& \#7 &	FL & FL & \#3 \\
& 0.160 & 0.116 & 0.139 & 0.549 & 0.072 \\
20 & FS	& FS & FL & log-D & FL \\
& 0.026	& 0.032	& 0.049 & 0.105 & 0.154 \\
21 & \#8 & \#5 & FS & \#1 & FL \\
& 0.146 & 0.096 & 0.017 & 0.043 & 0.191 \\
22 & FS	& \#4 &	FL & FS	& FS \\
& 0.024 & 0.079 & 0.173 & 0.026 & 0.012 \\
23 & log-D & \#6 & log-D & \#9 & \#2 \\
& 0.145	& 0.110 & 0.146 & 0.162 & 0.048 \\
\hline
\end{tabular}
\end{table}
%
%
%
%
%
%
\begin{figure}[h]
\includegraphics[trim=2cm 14cm 1cm 1cm,clip,width=8cm]{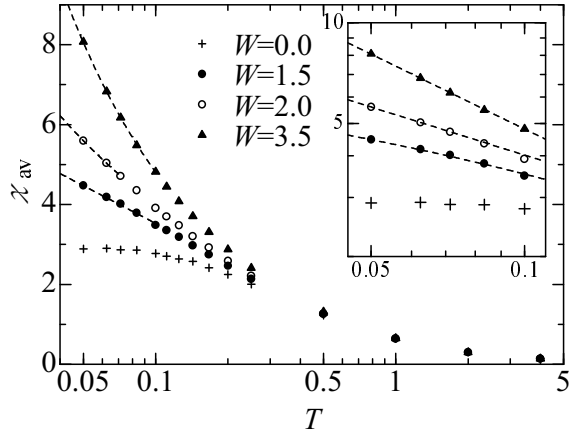}
\caption{Average susceptibility for $W = 1.5$, $2.0$, and $3.5$. 
The broken lines are fitted by the least-squares method. 
For W=1.5, 2.0, and 3.5, $\chi_{\rm av}(T) = -1.36 \, \log T +0.397$, $1.34 \, T^{-0.477}$, and $0.867 \, T^{-0.744}$, respectively. 
Inset: $\log \chi_{\rm av}(T)$ versus $\log T$. 
}
\label{Fig2}
\end{figure}
%
%
%
%
$\chi_{\rm av}(T) \sim T^{- \alpha}$ with $\alpha \sim 0.477$ ($W=2.0$) and $\alpha \sim 0.744$ ($W=3.5$). 
Therefore, in 2D disordered electron systems with dilute magnetic impurities, the observable susceptibility shows a non-Fermi-liquid behavior in low temperature.

Numerical computation was partly carried out at Supercomputer Center, the Institute for Solid State Physics, University of Tokyo. 
This work was partly supported by a Grant-in-Aid for Scientific Research from the Ministry of Education, Culture, Sports, Science, and Technology, Japan. 


\end{document}